**Fractal dimensions of umbral and penumbral regions of sunspots**

Rajkumar, B.[1] Haque, S.[1] Hrudey, W. [2]

[1]Department of Physics, University of the West Indies, St Augustine, Trinidad, W.I.

[2] William Hrudey Observatory, University College of Cayman Islands, Grand Cayman

**Abstract:**

The images of sunspots in sixteen active regions taken at the UCCI Observatory in Grand Cayman during June - November 2015, were used to determine their fractal dimensions using the perimeter-area method for the umbral and the penumbral region. Scale free fractal dimensions of $2.09 \pm 0.42$ and $1.72 \pm 0.4$ were found respectively. This value was higher than the value determined by Chumak and Chumak (1996) who used a similar method but for the penumbral region only for their sample set. There is a positive correlation $r = 0.58$ between umbral and the penumbral fractal dimensions for the specific sunspots. Furthermore, a time series analysis was done similarly on eight images of AR 12403, from 21[st] August 2015 – 28[th] August 2015 taken from the Debrecen Photoheliographic Data (DPD). The correlation $r = 0.623$ between the umbral and penumbral fractal dimensions in the time series indicating that the complexity in morphology indicated by the fractal dimension between the umbra and penumbra followed each other in time as well.

**Keywords:** active regions; sunspots; fractal dimensions; umbra; penumbra

# 1. Introduction:

Fractals were popularised by Mandelbrot (1983) and since then have become very useful tools for the quantification of irregular shapes and dynamical phenomena in nature. The wide range of their applications include use in diverse fields such as examining the growth of cancer cells (Losa, 2012) to the behaviour of stock exchanges (Panas 2001). Their applications are extensive and Morse (1985) even applied the use of fractals on vegetation and the body sizes of insects that can live on it. They found that very tiny insects occupy those with high fractal values or complexity. Practical applications of fractal dimensions have aided in determining the measure of toughness as a function of roughness in concrete (Issa *et al*., 2003). Wherever, there is evidence of irregular and complex patterns in nature that may be static, it has been a useful tool, but its use also extends to dynamical phenomena. Fractals can help determine if a system has turbulence and is chaotic (Mazzi and Vassilicos, 2004). Bershadskii (1990) showed that large scale fractal structure in laboratory turbulence, the ocean and the clustering of galaxies can have a common percolation nature which can be approximated by the fractal dimension of the order of about 4/3.

When it comes to understanding solar phenomena, it is understood that solar magnetic activity lies at the heart of many solar features that are observed which include faculae, sunspots, solar flares, prominences and coronal mass ejections. It is modelled by dynamic chaotic system (Lawrence *et al*., 1995; Bofetta *et al*., 1999). The appearance, growth and disappearance of sunspots have been found to be connected to variations in the magnetic field of the Sun (Solanki, 2003). It has been shown by Deng *et al*. (2016a) that there are relationships between number of faculae, sunspot counts and sunspot areas and solar flare activity. Therefore, the behavior of any single phenomena is related to the behaviour of the other features as they share a common underlying magnetic dynamo process driving them. The physical distribution of such features on the solar disk is also not random, nor symmetrical between northern and southern solar hemispheres. Deng *et al*. (2016a, 2016b) suggest that the nonlinear coupling of the polar magnetic fields with strong active-region fields produces the complexity and the relationships between the polar faculae and sunspot numbers and areas. They used a time series to investigate the fractal and chaotic properties of high and low-latitude solar activity by use of the fractal dimension, Hurst exponent and Lyapunov exponent and thereby were able to establish a predictability timescales in years for these three measures. They determined values of the fractal dimension as approximately 1.2 and the Hurst exponent of approximately 0.8 showed long range persistence (Deng *et al.* 2016a). The fractal dimension for sunspots determined at a point in time differs from the value determined through time series, that is, spatial and temporal determinations yield different values (Georgoulis 2005). In this paper, we determine the scale-free fractal dimension d, from images of spatially distributed active regions.

Deng (2016c) further analyzes the magnetic complexity and multi-fractal behavior of solar Hα flare activity and finds that there is a long range correlation due to the multi-fractal behaviour and that the solar flare activity is most irregular in the northern hemisphere compared to elsewhere in the solar disk. Solar flares and sunspot activity has been strongly correlated (Yan and Qu 2007; Yan *et al*. 2009) as the stored magnetic energy in active regions suddenly gets released as kinetic energy of the particles, radiation and plasma flow and heat (Fletcher *et al*., 2011). The process of the transformation of the stored energy in the active region to solar flares however is not well understood. It is therefore important to understand the fractal behavior of solar flares as well to gain an insight into active regions on the Sun. The findings by Deng (2016c) on the solar flares follow the non symmetrical behavior of active regions in the different hemispheres of the Sun and is consistent with it.

Deng *et al*. (2016b) have also investigated constraints for the solar dynamo models by investigating the hemispheric interconnection of solar activity phenomena for sunspot areas during Solar Cycless 9 – 24. They note that sunspot areas are suitable indicators of solar magnetic activity as well as giving an insight into the growth and decay of sunspots (Feng *et al*., 2014). They found that the sunspot areas have greater physical significance than the sunspot numbers since there is a linear relationship between sunspot areas and its total magnetic fluxes (Preminger and Walton 2006). It is noted that sunspot areas as a measure of solar magnetic activity are more suitable than sunspot numbers because there are challenges in the measurement and identification of the smallest spots (de Toma *et al*., 2013). The study in this paper uses area-perimeter relationship to determine the fractal dimension.

The monthly sunspot group numbers have been found to vary for the northern and southern hemispheres of the solar disk (Deng *et al*., 2013) over different eras. The monthly sunspot group numbers in the northern hemisphere preceded those in the southern hemisphere during 1874-1927, with a subsequent shift to the southern hemisphere leading the northern hemisphere during 1928-1964 and Deng *et al*. (2013) note that currently we are in the phase where the monthly sunspot numbers in the northern hemisphere lead the southern hemisphere counts. Therefore in any study with sunspots, it is important to highlight their location. The periodicity of solar activity in both hemispheres have not been found to be identical (Deng *et al*., 2014). Our sunspot sample can be classified as being located in the low-latitude band from -18.4 to 17.2 degrees latitude.

Qin (1994) performed a fractal study on sunspot relative numbers and determined a fractal dimension D = 2.8 ± 0.1 using data from January 1850 to May 1992. This study was extended by Qin (1996) who employed fractal dimensions to assist in prediction of monthly sunspot numbers using the theory of nonlinear dynamical systems. Greenkorn (2009) examined the nonlinear analysis of the daily sunspot number for each of cycles 10 to 23 to determine if the convective turbulence was stochastic or chaotic. He found that there was stochastic behaviour for cycles 10 to 19, transitioning to chaotic behaviour for cycles 20, 21, 22, and 23, having implication for the scale of turbulence. However, Price *et al*. (1992) found that using raw monthly sunspot number data for a 22 year period (monthly mean Wolf sunspot numbers), shows no evidence that the sunspot numbers are generated by a low-dimensional deterministic nonlinear process. More recently, Gayathri and Selvaraj (2010) also used sunspot numbers as a measure of solar activity using fractals. For the period 1990 to 2004, they found that the average fractal dimension for periods of 10 days or less was around 1.43 but changed to 1.72 for periods longer than 10 days. Shenshi *et al*. (1999) analyzed the dynamic behaviour for the monthly mean variations of the sunspot relative number from January 1891 to December 1996 and found a fractal dimension of $3.3 \pm 0.2$. This is the highest fractal dimension encountered in such studies to date. Fractals application to sunspot numbers have been one of the most common application of this tool to solar activity.

Watari (1995) studied the fractal dimensions of solar activity temporally using daily solar indices like the sunspot number, 10.7 cm radio flux, coronal emission and the total solar irradiance from the NIMBUS-7/ERB *(Earth Radiation Budget)*. The author found that the solar activity varies more irregularly for time scales that are longer than several days and shorter than several months and, that the yearly values of the fractal dimension did not change in concert with the Solar Cycle.

Zelenyi and Milovanov (1991) also presented a fractal model of sunspots obtaining expressions for the magnetic field distributions in sunspot umbrae and penumbrae. The model allowed a qualitatively explanation for the process of sunspot formation and, the morphology of spots with developed penumbras with the fractal dimension being 1.24 for developed sunspots.

Implication of turbulent structure by use of fractal analysis to sunspots was studied by Chumak and Chumak (1996) where they determined the fractal dimension of the solar sunspot umbra using cross-section area – outline length method which we have used. They found a value for the fractal dimension 1.35 which corresponded to turbulent structure in percolation theory. Chumak and Zhang (2003) further explored the question of whether total sunspot area and the total magnetic flux were proportional to each in ten solar Active Regions. While they found that some of the relationships satisfied simple power laws, no single power law for the area-flux correlation was common to all the Active Regions. Thus fractal examination showed that what some of the power laws found could not be justified inside the simple models of stationary magnetic flux tube aggregation. Chumak (2005) extended the study to include self-similar and self-affine structures on the sunspots and their magnetic fields. Solar X-ray flux was also studied and could be represented as well by fractal dimension.

In this paper we have followed the analysis of Chumak and Chumak (1996) but, applied it to the umbral as well as the penumbral region as well. Fractals can help to determine the nature of the sunspot morphology and in so doing can perhaps help in predicting their evolution from the stage of their formation to their disappearance. Much of the recent work has focused on the time series analysis to determine the multifractal dimension (Vertyagina and Kovslovskiy, 2013) while this paper focuses on the complexity of the actual structure of two aspects of the sunspots, the umbra and penumbra as possible indicators of its evolution. It has not been common practice to consider both the umbra and penumbra in studies of active regions. Since such features are dynamical phenomena and the morphology can clearly define these aspects common to most sunspots, it can be an indicator of its evolution. We also present some preliminary results on using this same method to determine the evolution of the fractal dimension temporally for both the umbral and penumbral regions for a set of eight images of AR 12403, from 21st August 2015 – 28th August 2015 obtained from the Debrecen Photoheliographic Data (DPD).

The paper is divided as follows. In Section 2, the detail of the digital solar imaging for the images used from University College of Cayman Islands (UCCI) Observatory is given.  In Section 3 we report on the data used. Section 4 reports on the results and analysis on the data. In Section 5, we discuss our findings and conclude in Section 6.

## 2. Digital Solar Imaging:

Although the UCCI Observatory has a dedicated 8" Solar Newtonian, the images used in this study were captured with a TS 115mm APO triplet refractor equipped with a 3x's TeleVue Barlow, Baader Herschel wedge, Baader Solar Continuum filter (540nm bandpass) and IR/UV cut filter.   Digital imaging devices included both the monochrome DMK41 (1280 × 960 pixels of 4.65um) and ZWO ASI120MM (1280 × 960 pixels of 3.75um).  SharpCap 2.6 was used for capture software in which 300 frames were captured as an AVI file.  Stacking was carried out with AutoStakkert2 software where 15% of the ranked best frames were stacked.  This process improves SNR and, the threshold is adjustable.  For the next level of processing, we used Registax 6 which provided a very effective "wavelet" function which improves image quality further.  Finally, for photo editing, we used PaintShop Pro 7 for the addition of false color and adjustment of brightness/contrast.  Magnification was not an issue since the parameters to be measured were ratios rather than absolute.

The measurement of the area and perimeter were determined using ImageJ[1]. It offers multiple graphics and analytical tools of which two allow for rapid and accurate measurement of area

---

[1] Imagej.nih.gov. 2016. "ImageJ." accessed 14, November.

and perimeter of a defined entity. The threshold is adjustable based on adjacent pixel gradients.

**3. Data**:

The image set contained 16 ARs taken between 16[th] June 2015 and 3[rd] November 2015. The images were inspected to ensure only good images were used. Good images were defined by two criteria (Solanki 2003).

i.   Images must have been taken when the ARs were near the centre of the solar disc so as to avoid distortions caused by the Wilson Effect.

ii.  Images must contain the highest level of contrast available.

**For this study the area and perimeter of the sunspot umbra and penumbra in each active region were determined. Active regions were defined as clusters of sunspots and pores which were clearly connected and given an official number by NOAA[2]. For each AR, measurements were only taken for well-defined sunspots containing a clear umbra and penumbra. Pores which do not contain a penumbra were not included in the measurements. All measurements were done using the ImageJ software. The wand (tracing) tool was used to outline the umbrae (as seen in Figures 1b, 2b, and 3b) after which the measure tool was used to determine the areas and perimeters. The area is determined by counting all the pixels within the enclosed area while the perimeter counts the pixels used to form the outline. This was also done outlining the penumbrae**

as seen in Figures 1c, 2c and 3c. It should be noted that the penumbra outline enclosed both the umbra and penumbra. In active regions which contained multiple spots, all the spots containing a clear umbra and penumbra were outlined and measured as one entity.

Figures 1a to 3a show examples of some of the images used. Figures 1b to 3b show how the umbra was selected and Figures 1c to 3c show the selection of the penumbra respectively.

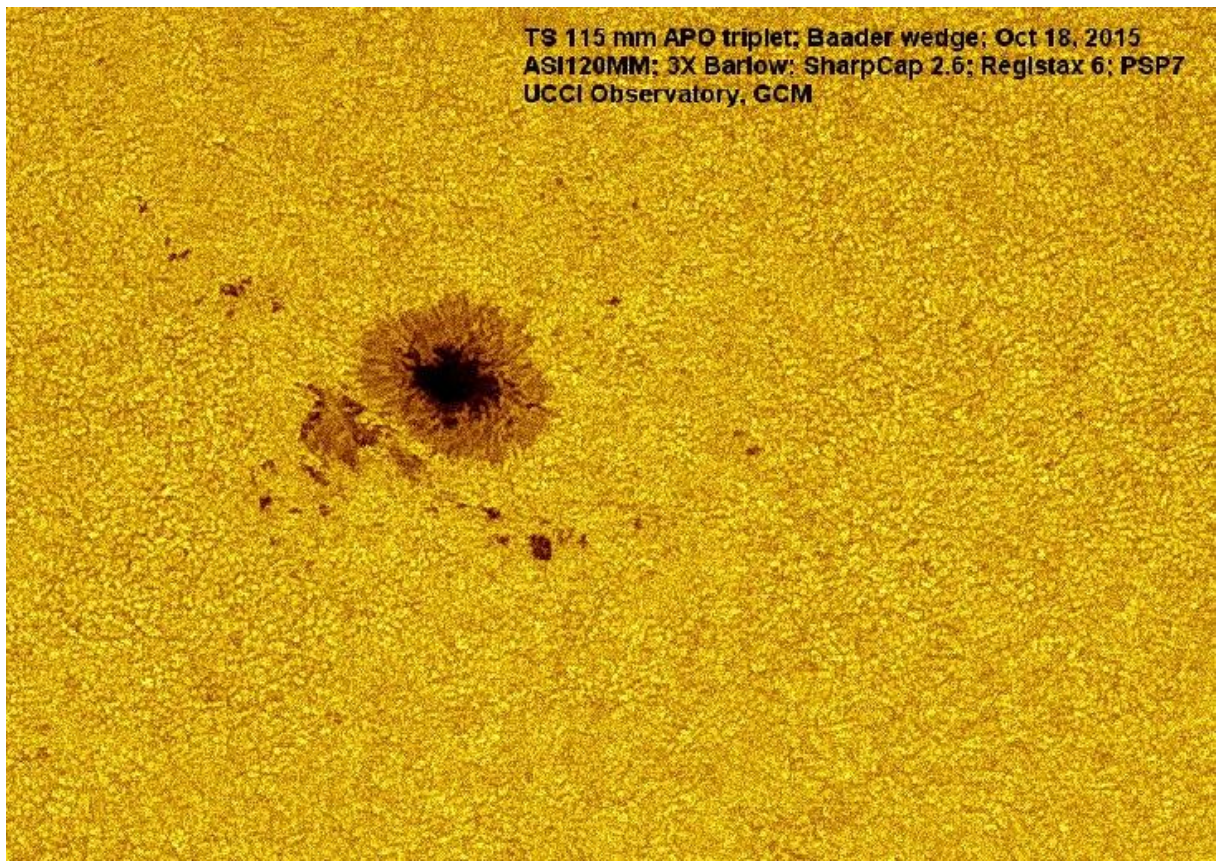

**Figure 1a**. Image of sunspot in AR 12434

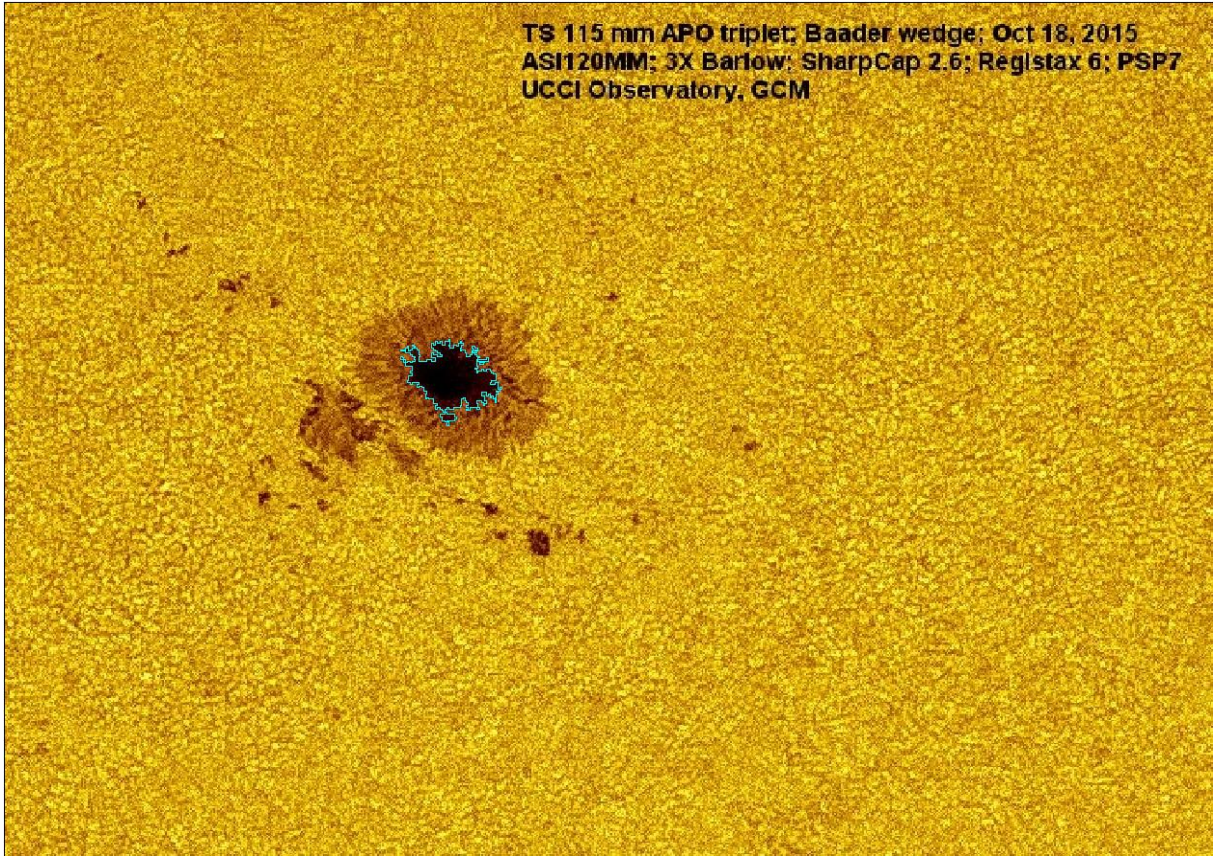

**Figure 1b.** Image of sunspot in AR 12434 with the umbra selected.

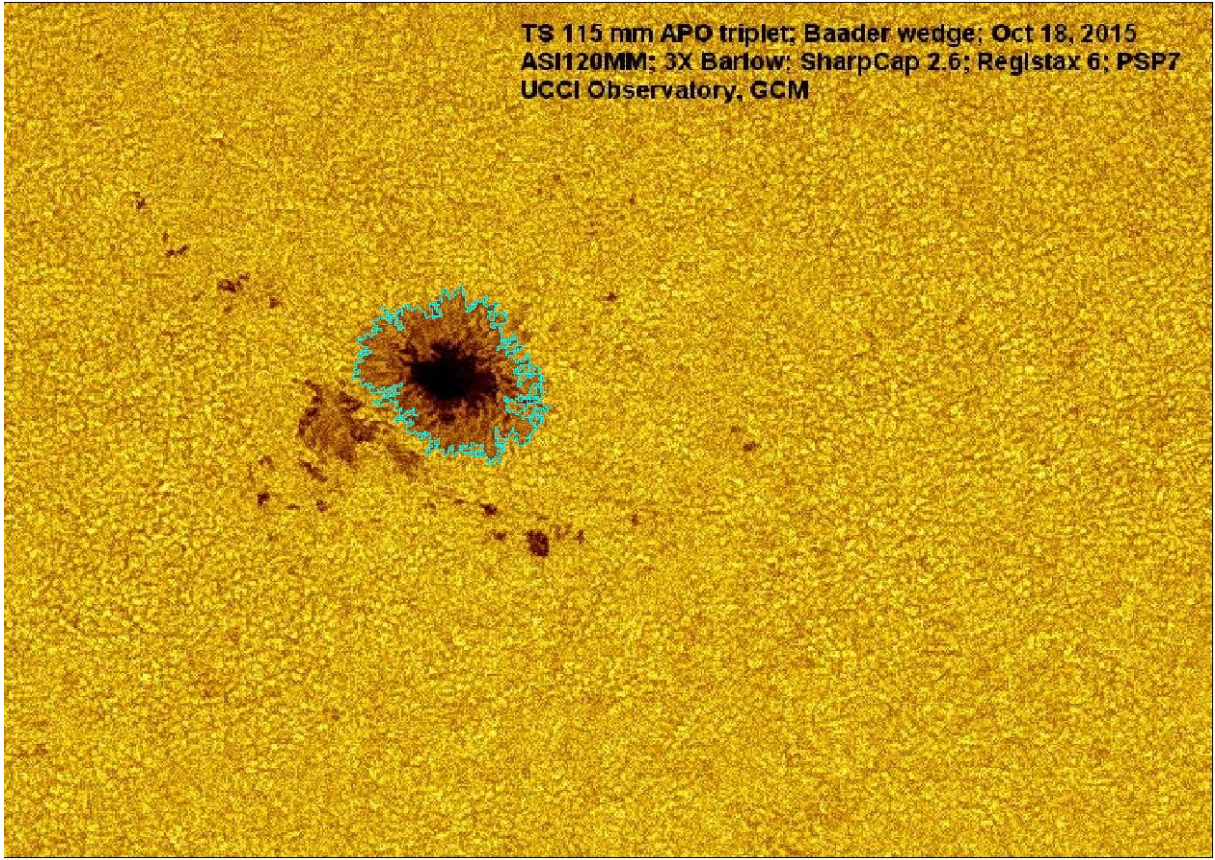

**Figure 1c**. Image of sunspot in AR 12434 with the penumbra selected.

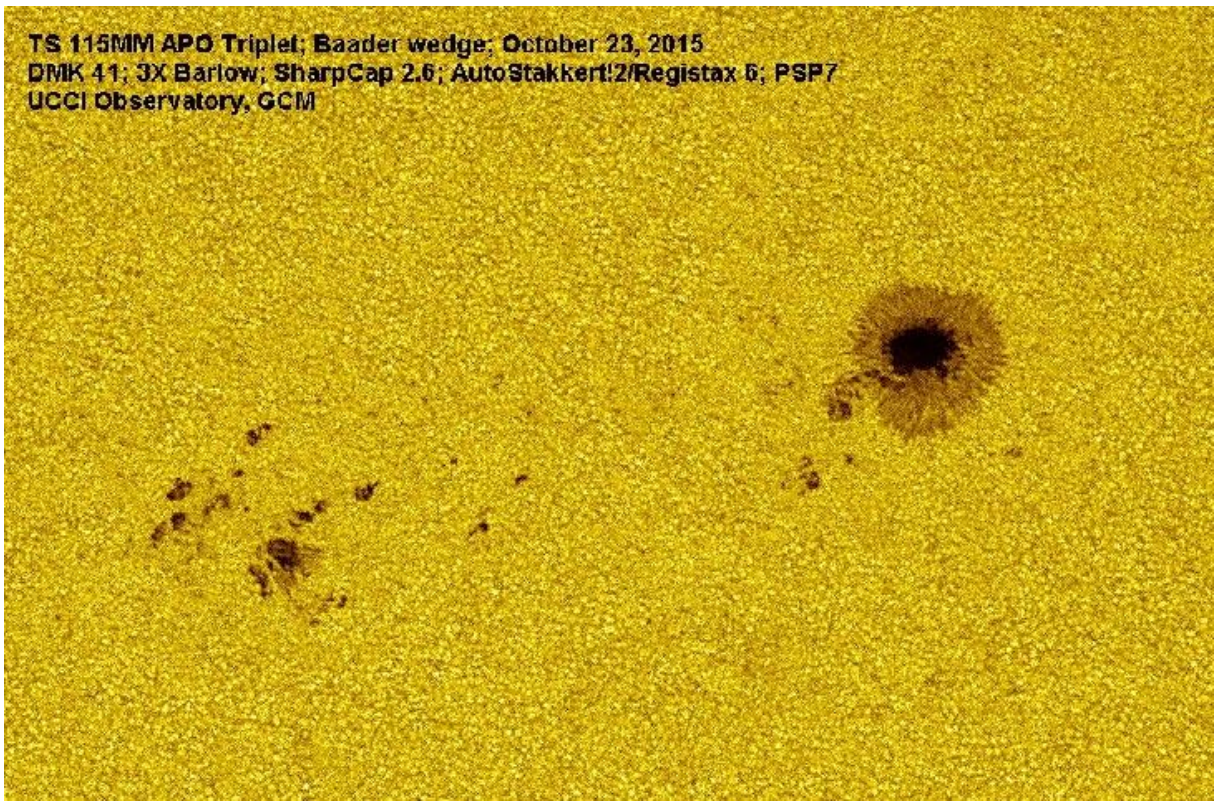

**Figure 2a.** Image of sunspot in AR 12436.

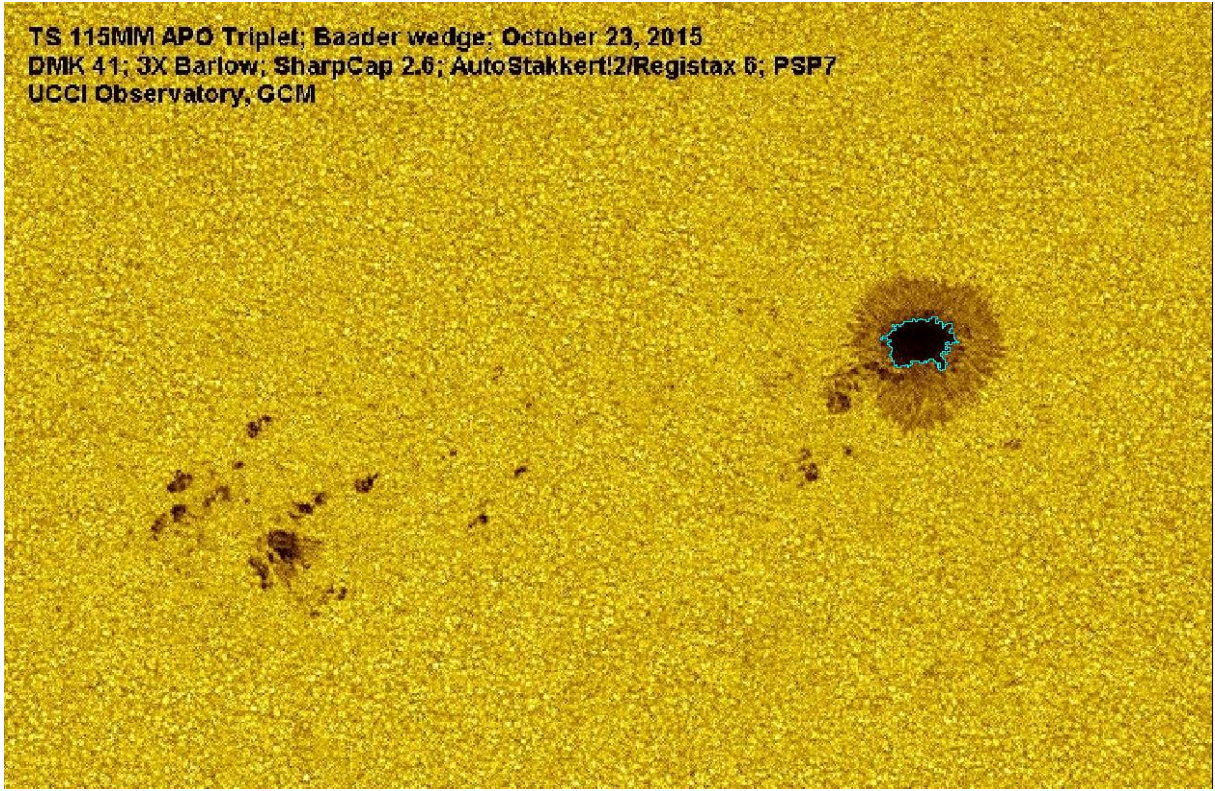

**Figure 2b.** Image of sunspot in AR 12436 with umbra selected.

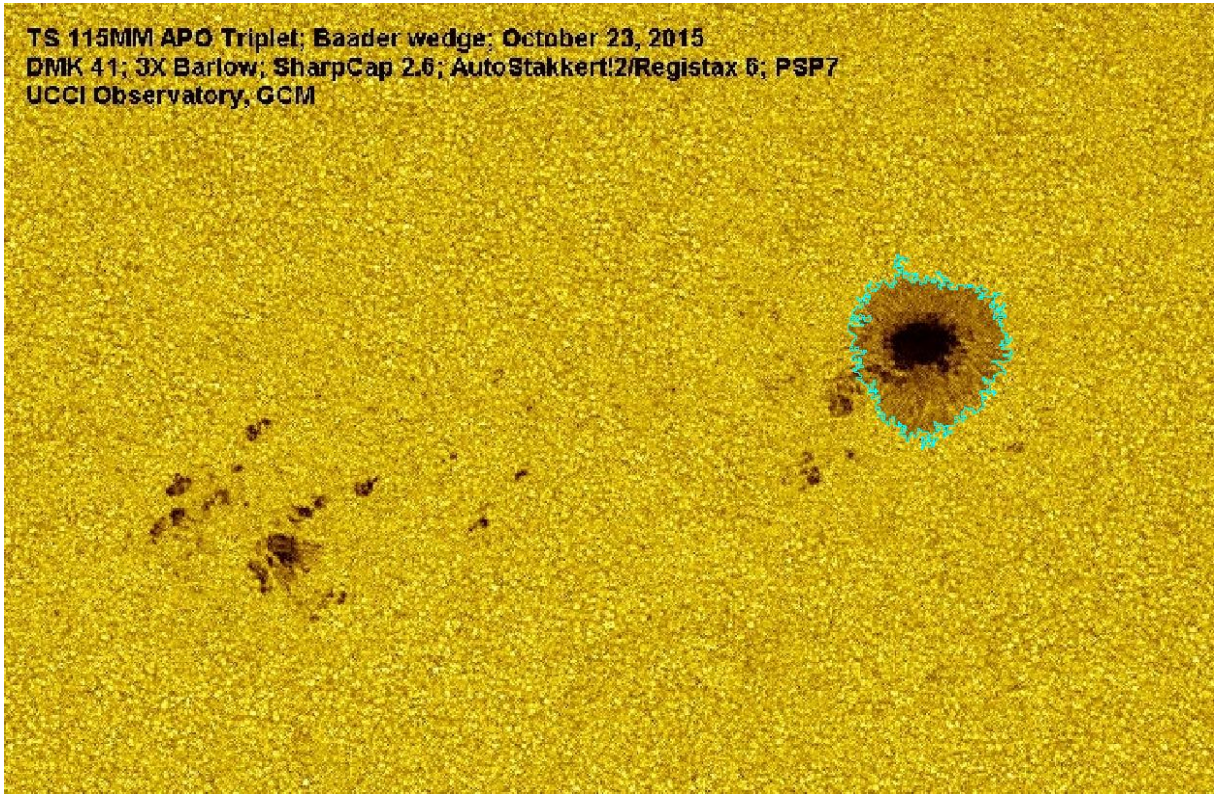

**Figure 2c.** Image of sunspot in AR 12436 with penumbra selected.

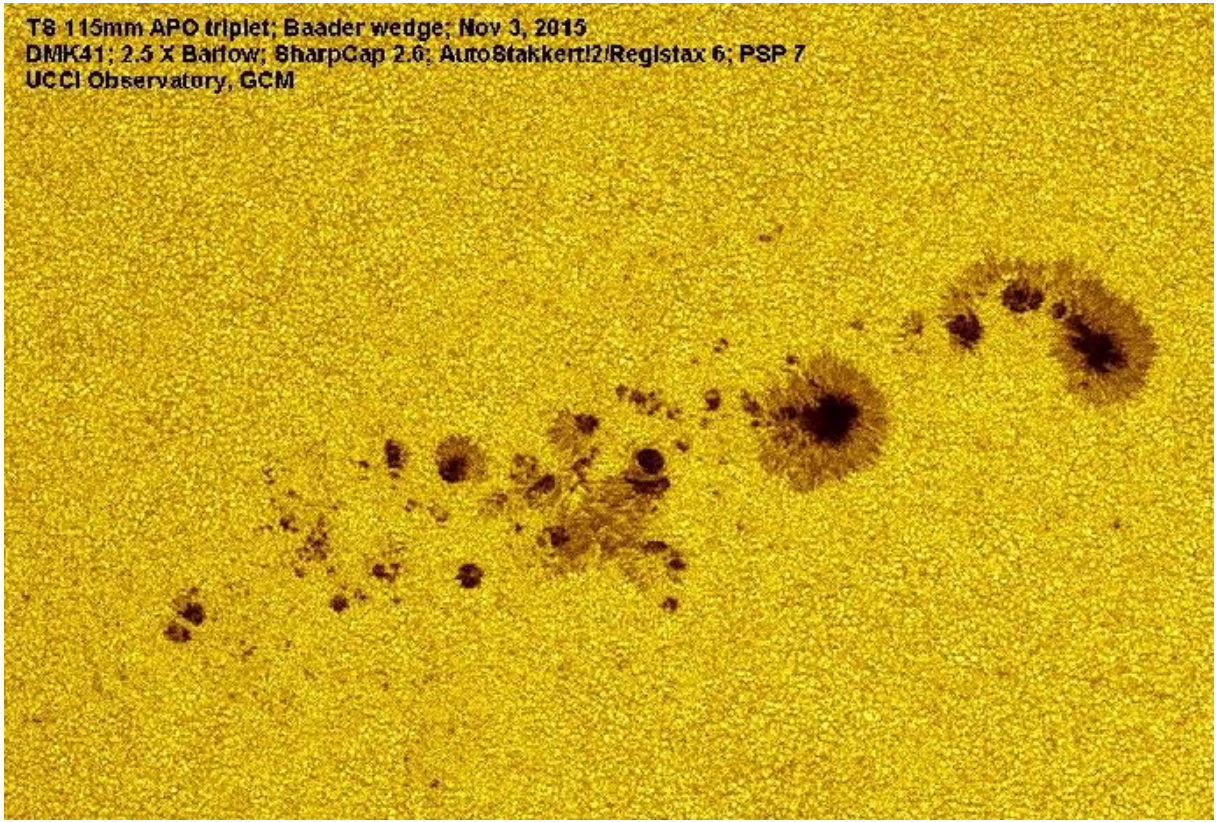

**Figure 3a.** Image of sunspot AR 12443.

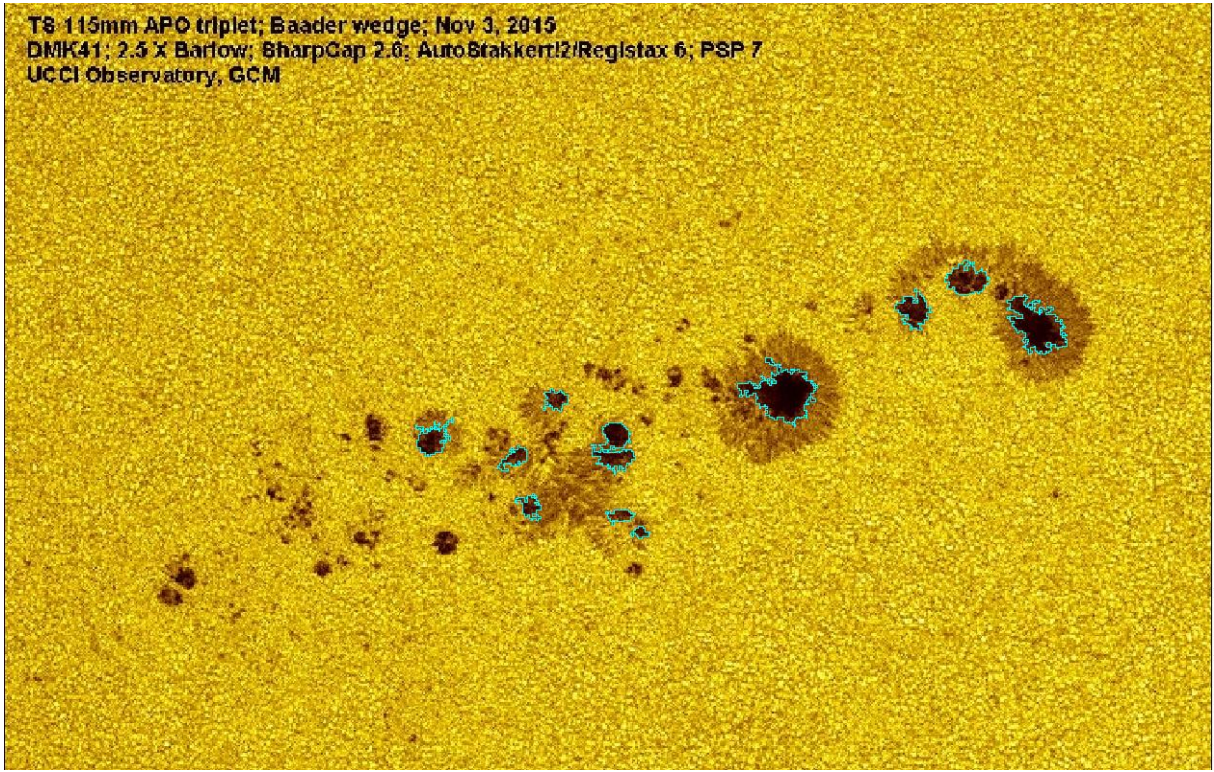

**Figure 3b.** Image of sunspots in AR 12443 with the umbrae selected.

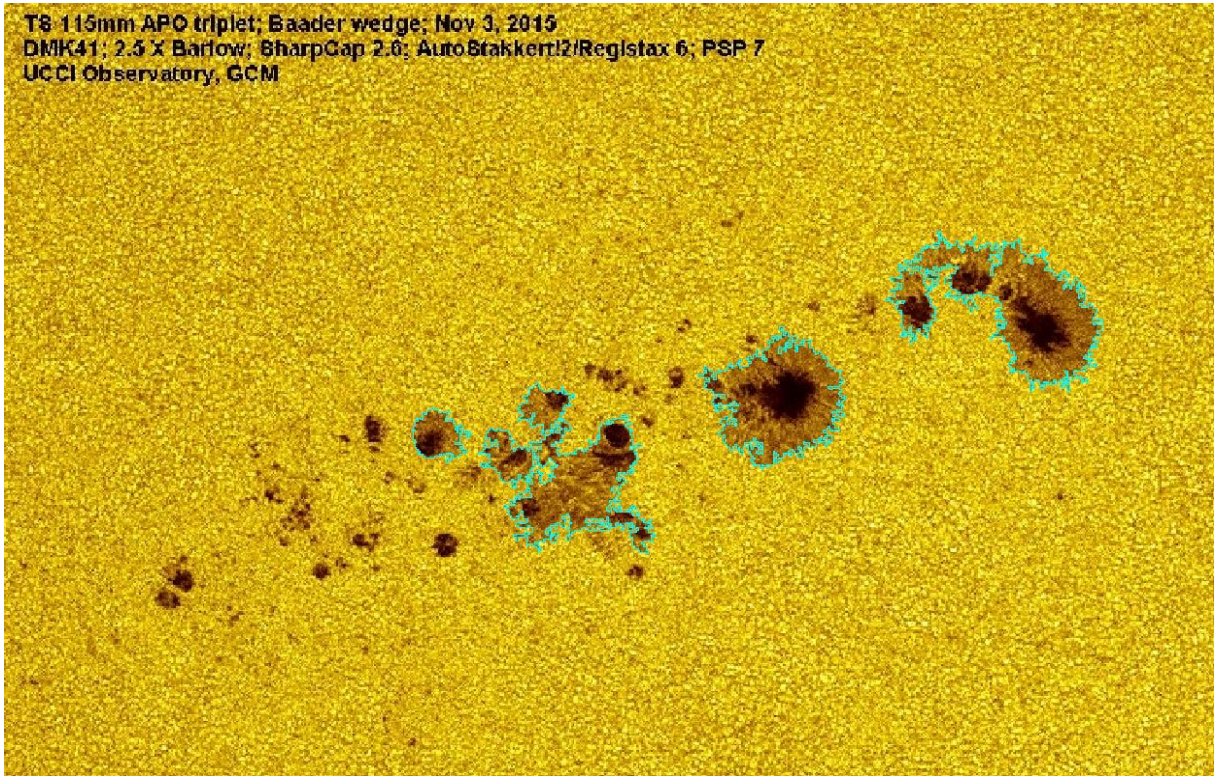

**Figure 3c.** Image of sunspots in AR 12443 with the penumbrae selected.

## 4. Results and Analysis:

To determine the fractal dimensions of the sunspot's umbrae and penumbrae, the area-perimeter relation as defined by Chumak and Chumak (1996) was used.

The area of the sunspot umbra/penumbra, S, is related to the perimeter of the sunspot umbra/penumbra, L, as follows:

$$S \sim L^q,$$

$$q = \frac{Log\ S}{Log\ L},$$

$q$ is related to the fractal dimension, $d$,

$$d = \frac{2}{q},$$

if the area is fractally dependant on the perimeter.

Table 1 gives the measurements for the umbrae and penumbrae of the sunspots in the 16 Active Regions selected for analysis with the timing and date of data acquisition as well as the latitude where that was available.

| AR # | Date | Time | Latitude | Umbra | | | | Penumbra | | | |
|---|---|---|---|---|---|---|---|---|---|---|---|
| | | | | Area-S (pixels) | Perimeter-L (pixels) | log S | log L | Area-S (pixels) | Perimeter-L (pixels) | log S | log L |
| 12367 | 16/06/2015 | 18:04:11 | - | 2706 | 972.07 | 3.43 | 2.99 | 10639 | 2835.23 | 4.03 | 3.45 |
| 12371 | 22/06/2015 | 15:12:34 | 12.16 | 8050 | 2392.11 | 3.91 | 3.38 | 36698 | 5941.49 | 4.56 | 3.77 |
| 12373 | 03/07/2015 | 17:22:13 | - | 1854 | 860.76 | 3.27 | 2.93 | 5943 | 1475.29 | 3.77 | 3.17 |
| 12381 | 10/07/2015 | 15:01:38 | 13.51 | 3595 | 814.62 | 3.56 | 2.91 | 14569 | 2119.88 | 4.16 | 3.33 |
| 12384 | 14/07/2015 | 15:31:42 | -18.38 | 1559 | 297.14 | 3.19 | 2.47 | 6597 | 943.62 | 3.82 | 2.97 |
| 12386 | 20/07/2015 | 14:52:04 | - | 1409 | 403.87 | 3.15 | 2.61 | 6235 | 1464.71 | 3.79 | 3.17 |
| 12387 | 20/07/2015 | 15:10:08 | - | 1865 | 787.67 | 3.27 | 2.90 | 3962 | 1367.29 | 3.60 | 3.14 |
| 12394 | 06/08/2015 | 14:11:25 | 12.38 | 2708 | 553.11 | 3.43 | 2.74 | 11021 | 1859.11 | 4.04 | 3.27 |
| 12396 | 10/08/2015 | 13:40:19 | -17.67 | 17702 | 5187.56 | 4.25 | 3.71 | 64115 | 9993.90 | 4.81 | 4.00 |
| 12400 | 14/08/2015 | 14:15:26 | 17.21 | 1369 | 420.48 | 3.14 | 2.62 | 3991 | 1448.97 | 3.60 | 3.16 |
| 12401 | 18/08/2015 | 13:32:07 | - | 1365 | 473.60 | 3.14 | 2.68 | 5422 | 1682.96 | 3.73 | 3.23 |
| 12403 | 22/08/2015 | 13:29:06 | -15.34 | 14199 | 3647.31 | 4.15 | 3.56 | 78465 | 9805.05 | 4.89 | 3.99 |
| 12418 | 20/09/2015 | 15:35:39 | -15.54 | 2546 | 615.42 | 3.41 | 2.79 | 11404 | 1374.26 | 4.06 | 3.14 |
| 12434 | 18/10/2015 | 15:18:49 | -9.87 | 1938 | 782.18 | 3.29 | 2.89 | 10069 | 1758.72 | 4.00 | 3.25 |
| 12436 | 23/10/2015 | 15:39:39 | 8.74 | 1738 | 632.94 | 3.24 | 2.80 | 7357 | 1769.77 | 3.87 | 3.25 |
| 12443 | 03/11/2015 | 15:07:55 | 6.60 | 3016 | 1265.68 | 3.48 | 3.10 | 16376 | 4284.58 | 4.21 | 3.63 |

**Table 1**. Measurements for sunspot umbrae and penumbrae for active regions (ARs) with the date and time of image acquisition.

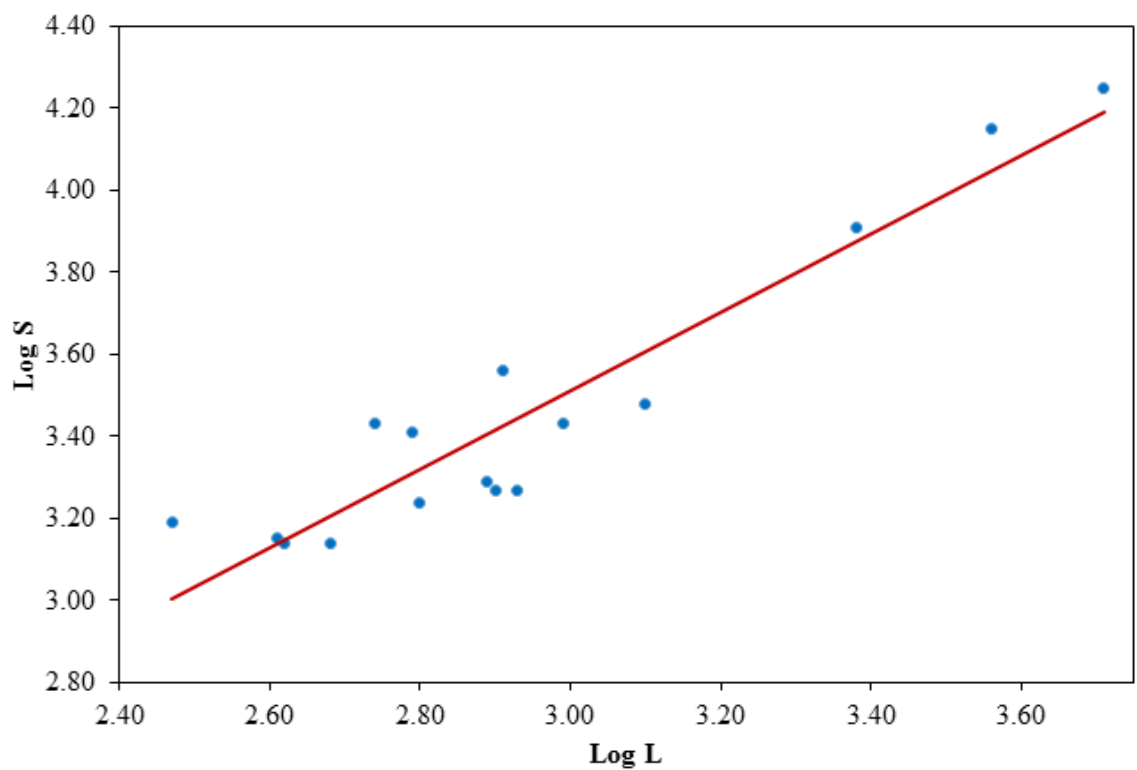

**Figure 4**. Determination of the fractal dimension *d* for umbrae from plots of area S and perimeter L.

Therefore, $q = 0.96 \pm 0.19$ and fractal dimension, $d = 2.09 \pm 0.42$ with Pearson correlation of 0.94 from Figure 4.

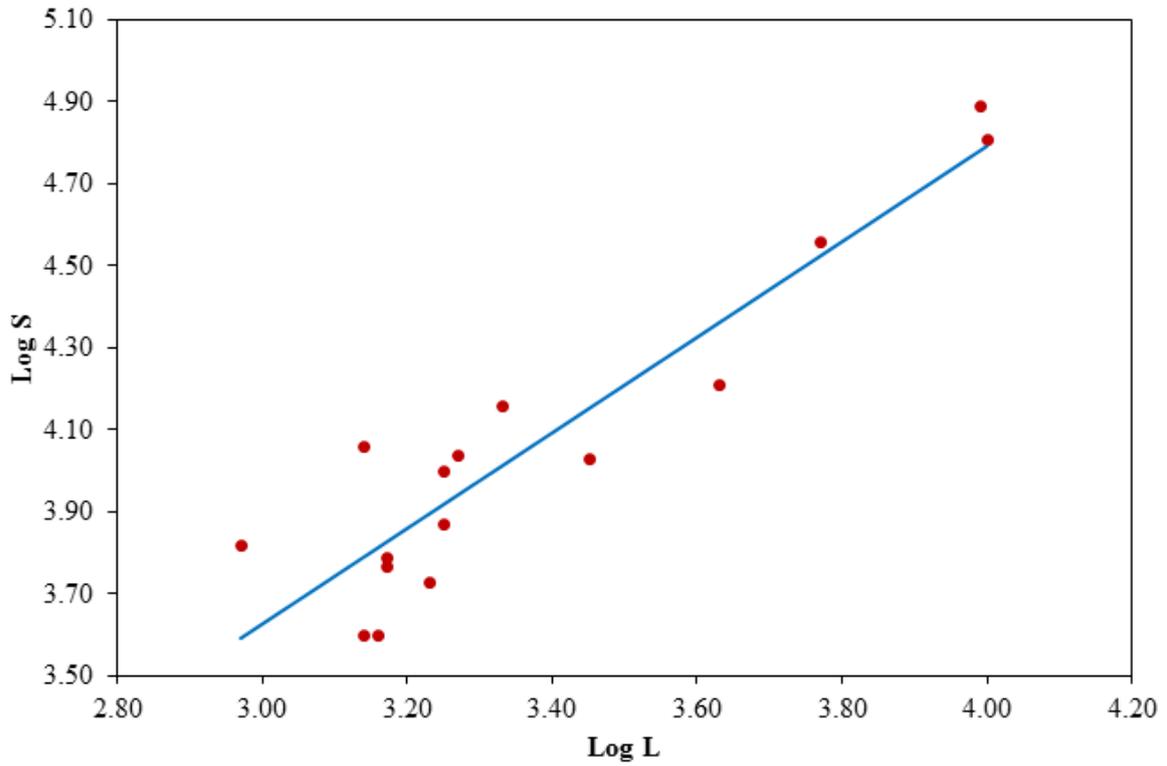

**Figure 5.** Determination of the fractal dimension d for penumbrae from plots of area S and perimeter L.

From Figure 5, therefore, $q = 1.16 \pm 0.27$ and fractal dimension, $d = 1.72 \pm 0.40$ with Pearson correlation of 0.927.

Table 2 lists the fractal dimensions of the umbra and the penumbra for the respective sunspots for the sunspots in the 16 active regions analyzed.

| AR # | Umbra | | Penumbra | |
|---|---|---|---|---|
| | q | d=2/q | q | d=2/q |
| 12367 | 1.15 | 1.74 | 1.17 | 1.71 |
| 12371 | 1.16 | 1.73 | 1.21 | 1.65 |
| 12373 | 1.11 | 1.80 | 1.19 | 1.68 |
| 12381 | 1.22 | 1.64 | 1.25 | 1.60 |
| 12384 | 1.29 | 1.55 | 1.28 | 1.56 |
| 12386 | 1.21 | 1.66 | 1.20 | 1.67 |
| 12387 | 1.13 | 1.77 | 1.15 | 1.74 |
| 12394 | 1.25 | 1.60 | 1.24 | 1.62 |
| 12396 | 1.14 | 1.75 | 1.20 | 1.66 |
| 12400 | 1.20 | 1.67 | 1.14 | 1.76 |
| 12401 | 1.17 | 1.71 | 1.16 | 1.73 |
| 12403 | 1.17 | 1.72 | 1.23 | 1.63 |
| 12418 | 1.22 | 1.64 | 1.29 | 1.55 |
| 12434 | 1.14 | 1.76 | 1.23 | 1.62 |
| 12436 | 1.16 | 1.73 | 1.19 | 1.68 |
| 12443 | 1.12 | 1.78 | 1.16 | 1.72 |

**Table 2.** Fractal dimensions of the umbra and penumbra for sunspots in active regions (ARs).

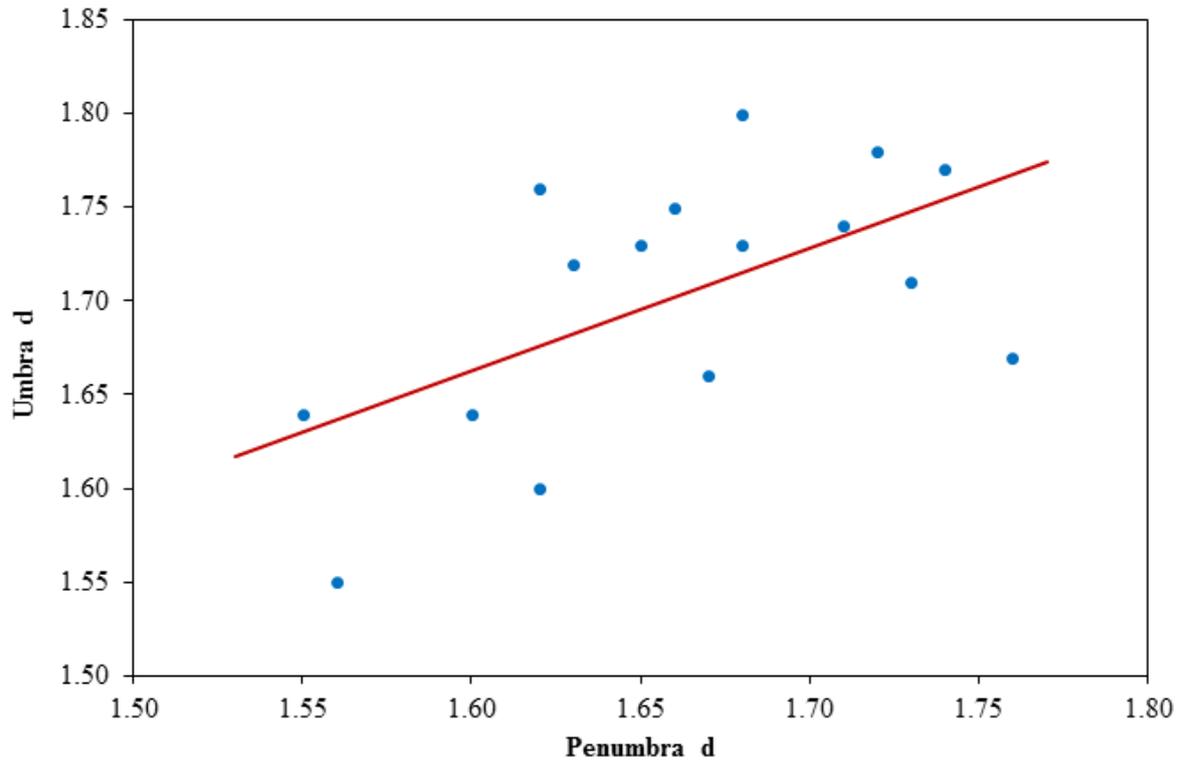

**Figure 6.** The fractal dimension of the umbra against the fractal dimension of the respective penumbra d for 16 ARs (*cf.* Table 2).

From Figure 6, it is found that there is a positive correlation between the umbra and penumbra fractal dimensions of 0.584.

A preliminary time series analysis was done on eight consecutive days following the same active region, which was emerging, through its disc passage. This data set comprises of eight images of AR 12403, from 21$^{st}$ August 2015 – 28$^{th}$ August 2015. The images are taken from the *Debrecen Photoheliographic Data* (DPD) (Baranyi, Győri, and Ludmány 2016) and is given in Table 3. **As can be seen in Figure 7, the central meridian passage occurs approximately in the middle of the period analyzed. The period is also done for eight days around the central meridian passage to ensure that distortion due to limb effects do not affect the data used.**

| Date | Time UTC | Latitude | Julian Date | Umbra | | | Penumbra | | |
|---|---|---|---|---|---|---|---|---|---|
| | | | | Area – S (pixels) | Perimeter – L (pixels) | d | Area – S (pixels) | Perimeter – L (pixels) | d |
| 21/08/2015 | 05:10:23 | -14.75 | 2457255.716 | 1289 | 522.839 | 1.75 | 12423 | 1832.085 | 1.59 |
| 22/08/2015 | 05:34:22 | -14.87 | 2457256.732 | 2716 | 676.345 | 1.65 | 13361 | 1682.584 | 1.56 |
| 23/08/2015 | 05:46:22 | -15.27 | 2457257.741 | 2595 | 748.9 | 1.68 | 20598 | 2451.713 | 1.57 |
| 24/08/2015 | 05:34:22 | -15.34 | 2457258.732 | 3267 | 797.329 | 1.65 | 25000 | 3327.94 | 1.60 |
| 25/08/2015 | 05:10:22 | -15.30 | 2457259.716 | 5615 | 1314.113 | 1.66 | 28862 | 2190.339 | 1.50 |
| 26/08/2015 | 05:58:22 | -15.19 | 2457260.749 | 3921 | 766.863 | 1.61 | 23385 | 1683.087 | 1.48 |
| 27/08/2015 | 05:46:22 | -15.07 | 2457261.741 | 2685 | 643.153 | 1.64 | 18564 | 1748.65 | 1.52 |
| 28/08/2015 | 05:10:22 | -14.88 | 2457262.716 | 2423 | 656.183 | 1.66 | 11303 | 1279.135 | 1.53 |

**Table 3.** Measurements for fractal dimensions for umbra and penumbra of AR 12403.

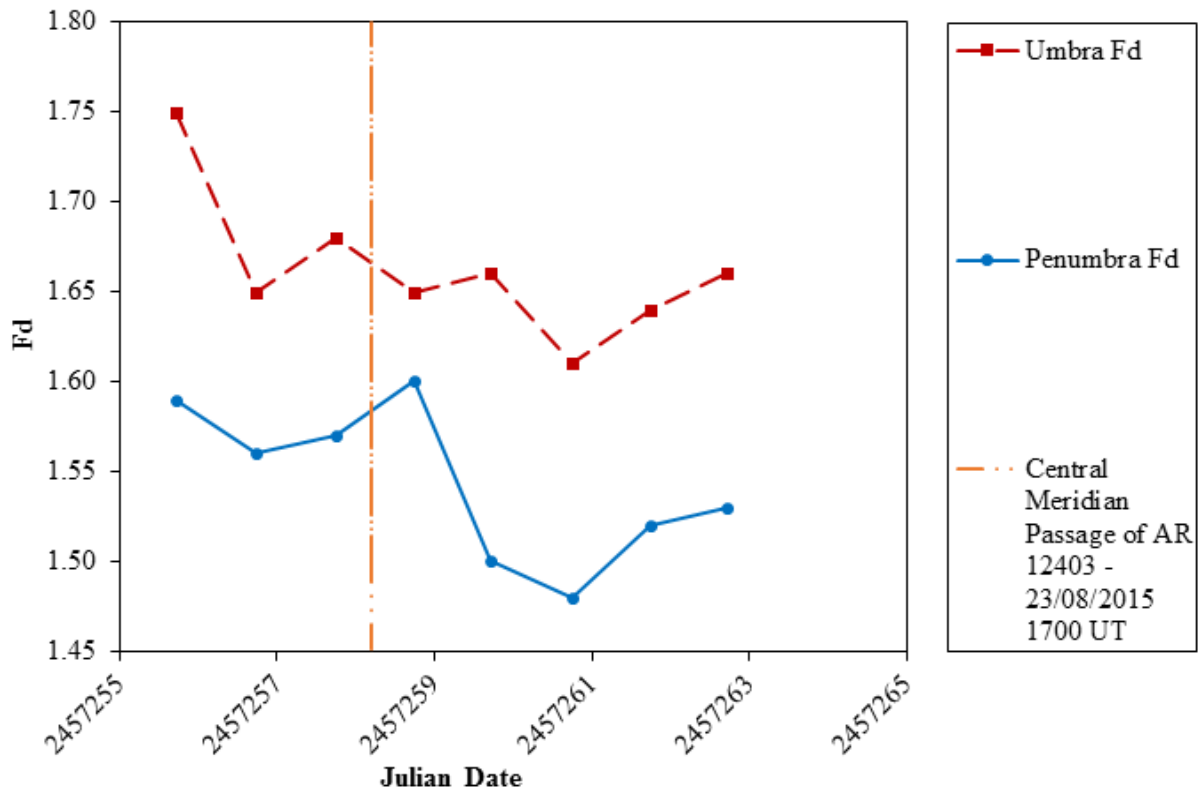

**Figure 7.** The time series of the fractal dimension for AR 12403 for the umbrae and penumbrae.

From Fig 7, it is found that there is a correlation coefficient r = 0.623 between the fractal dimension for the corresponding umbra and penumbra with time.

## 5. Discussion

As noted in the Introduction, fractal dimensions are very useful in a variety of dynamic and chaotic phenomena. While their importance in describing features associated with magnetic activity on the Sun such as faculae, sunspot numbers, coronal mass ejections and prominences has been highlighted, it is important to distinguish how the use and determination of fractal dimensions in this study varies from other uses of fractal dimensions for similar phenomena in other studies. Broadly, fractal dimensions can be used to quantify structure at a point in time or over a time i.e. spatially or temporally. This study has focused on the structure of sunspots spatially, similar to the analysis from Chumak and Chumak (1996) and unlike the studies by Deng *et al.* (2016c) which focused on the time series variations. Uritsky and Davila (2012) however studied the spatio-temporal evolution of unipolar and bipolar photospheric regions to analyze the fractal dimensions' multiscale behavior. Furthermore, fractal dimensions have also been determined for the distribution in the number of sunspots appearing on the solar disk (Deng, 2016a) showed that there is a correlation among the different measures of fractal dimensions of number of sunspots and time series determinations. These different determinations quantify the phenomena they measure which behave fractally and therefore obey a power law. Therefore, the distribution of sunspots, their temporal evolution, and their morphology all have shown fractal nature. This clearly indicates that magnetic activity giving rise to all of these determinations would be a scaling dynamical chaotic process. We used the correlation fractal dimension for our study with $q = 2$ which is widely used in studying morphology and is the scale free fractal dimension. This correlation dimension can be considered an empirical proxy for the correlation integral which yields the embedding dimension. Generalized correlation dimensions determines the multifractal nature within which the embedding dimension gives a measure of the correlation between different sets. Any discrepancies between these dimensions is therefore a measure of the clustering behavior (Uritsky and Davila 2012).

We applied the methodology of Chumak and Chumak (1996) to our 16 active regions in white-light, and we obtained a value for the penumbral fractal dimension of $1.72 \pm 0.4$, higher than their penumbral value of 1.35. We further determined the fractal dimension for the umbral region as well. This yielded a value of $2.09 \pm 0.42$. This indicates that the complexity of the structure within the umbral region is higher than that in the penumbral region. This can have implications as to how sunspots evolve. However, for such a study, the same spot has to be monitored over a period of time which is the current study underway by the authors. We report some preliminary results on this. For AR 12403, for a period of eight days from $21 - 28$ August 2015, the sunspot image from *Debrecen Solar Observatory (DSO)* was used to determine its umbral and penumbral fractal dimension. While this is preliminary result, it is interesting that umbral fractal dimensions are consistently higher than the penumbral values and also that they follow each other. There is a correlation coefficient between them of r = 0.623.

It is important to consider the effect of the resolution on the determination of the scale free fractal dimension. Was it possible that the difference in our findings could be due to the difference in the quality of the images as it is one of the features of fractals that the dimension can change with greater detail at the same scaling? However, the authors were unable to analyze the images of sunspots used in the study by Chumak and Chumak (1996). To test the effect of resolution, as a comparison, we computed the value of the fractal dimension for the unprocessed images from the UCCI Observatory used in this analysis and corresponding sunspot images obtained from the Debrecen Photoheliographic Data (DPD) (Baranyi, Győri, and Ludmány 2016). The resolution of the latter is reported to be as 1 arcsecond (Stanford.edu 2010). For the imaging system at the UCCI Observatory, we estimate the resolving power from the Dawe's limit as 1.01 arcsec while the diffraction limited resolution is 11.82 arcsec. The Table 4 below shows the 11 active regions used for this analysis on the impact on the fractal dimension of the resolution. The corresponding fractal dimensions from the three different datasets for a subset of 11 ARs that were available for the three different datasets is shown in Table 5.

| AR # | Date | Time UTC | Latitude |
|---|---|---|---|
| 12367 | 16/06/2015 | 18:04:11 | - |
| 12371 | 22/06/2015 | 15:12:34 | 12.16 |
| 12373 | 03/07/2015 | 17:22:13 | - |
| 12381 | 10/07/2015 | 15:01:38 | 13.51 |
| 12384 | 14/07/2015 | 15:31:42 | -18.38 |
| 12386 | 20/07/2015 | 14:52:04 | - |
| 12387 | 20/07/2015 | 15:10:08 | - |
| 12394 | 06/08/2015 | 14:11:25 | 12.38 |
| 12396 | 10/08/2015 | 13:40:19 | -17.67 |
| 12400 | 14/08/2015 | 14:15:26 | 17.21 |
| 12401 | 18/08/2015 | 13:32:07 | - |
| 12403 | 22/08/2015 | 13:29:06 | -15.34 |
| 12418 | 20/09/2015 | 15:35:39 | -15.54 |
| 12434 | 18/10/2015 | 15:18:49 | -9.87 |
| 12436 | 23/10/2015 | 15:39:39 | 8.74 |
| 12443 | 03/11/2015 | 15:07:55 | 6.60 |

**Table 4.** Active regions used for the resolution analysis from three datasets.

| | Raw (d) | Processed (d) | SDO (d) |
|---|---|---|---|
| Umbra | 2.06 ± 0.35 | 2.09 ± 0.33 | 2.17 ± 0.31 |
| Penumbra | 1.76 ± 0.64 | 1.87 ± 0.77 | 1.82 ± 0.37 |

**Table 5.** Fractal dimensions, d determined for the three data sets with different spatial resolutions.

Table 5 shows that the fractal dimension is not very sensitive to spatial resolution for the active regions. This is consistent with the findings of Georgoulis (2012) who determined that the multiscale parameters of active regions depended sensitively on resolution and observational characteristics rather than the scale free fractal dimension, which is what we have investigated. They determined this by utilizing three distinctly different spatial resolutions of the magnetograms with linear pixel sizes being 0.158 arcsec (highest resolution), 0.605 arcsec, and 1.98 arcsec (lowest resolution) to determine the fractal and multifractal dimension. They found that for 1.98 arcsec resolution the fractal dimension was 1.41, and for 0.605 arcsec it was 1.43 and the highest resolution yielded 1.54, all within error margins to be considered similar to our findings. Georgoulis (2012) notes that the values of the scale-free fractal dimension $d$, are fairly consistent, despite the widely different spatial resolution and the different instruments. He suggests that the reason for this might be fractal dimension qualitatively highlights the morphological complexity of the studied self-similar structure that is being reflected adequately on seeing-free *(Hinode* SOT/SP and SOHO/MDI) magnetograms largely regardless of spatial resolution and magnetic flux content.

An interesting study which yielded similar fractal dimensions as our study in a single system was reported by Weitz *et al*. (1985) on the limits of the fractal dimension for irreversible kinetic aggregation of gold colloids. While gold colloids seem to have nothing in common with sunspots, the former deals with the kinetic growth processes resulting in clusters. It is therefore possible that there may be similar kinetic growth processes leading to similar morphology for both aggregates and sunspots. As our images show, a perfectly spherical isolated sunspot is an extremely rare event compared to what can be generated by computer models (Rempel, 2011). The actual sunspot images appear quite similar on visual inspection to aggregates in the gold colloids. While Weitz *et al*. (1985) applied the fractal analysis to aqueous colloid systems, it is possible that the fluid dynamics of compressible flow dealing with the formation of the sunspots could be similar.

While there are several ways to determine the fractal dimension of a geometrical structure, it should be noted that the method used by Weitz *et al*. (1985) for the determination of the fractal dimension was through the determination of interparticle interactions and mass as a function of radius. The values they obtained indicate whether there is dominance of diffusion-limited kinetics or reaction-limited kinetics. This they controlled by the addition of pyridine to the gold colloids. By analogy, is it possible that the umbral and penumbral values of the fractal dimensions are reflective of similar processes occurring? Obviously the penumbral region is larger than the umbral region for a sunspot, and interestingly, the study on the gold colloids found that the diffusion-limited kinetics yields a larger cluster aggregate than the reaction-limited kinetics. The slow aggregation process yields the value of fractal dimension of 2.01, for $E > kT$ when determined from the mass of the clusters as a function of their radius. The diffusion-limited aggregates were determined to have a fractal value of 1.77, for $E << kT$, with both regimes being described by power laws.

By extension of the analogy to the results of this study, it is possible that the formation of the umbra and penumbra depend on energy and therefore temperature dynamics, however, on the macroscopic scale it should be noted that the darker regions of the umbra represent cooler temperatures than the brighter regions such as the penumbra.  The penumbra represents the transition of the temperature gradient to the ambient temperature of the solar surface and the end of the region of influence of magnetic activity contained within the active region.

Recent models of the sunspot formation, for example, Jaegglie (2012), hypothesized that ionized hydrogen becomes molecular hydrogen due to the reduced temperature in a sunspot, due to the trapping by the magnetic fields. We propose that the gold colloid analysis can be used as an analog model, since the addition of pyridine serves to displace the charged ions from the surface of the colloid even as the ionized hydrogen becomes molecular hydrogen according to Jaegglie (2012).

## 7. Conclusions

We have determined that the umbral and penumbral regions of the sunspots have different fractal dimensions from the analysis of sunspots in 16 active regions using the method of Chumak and Chumak (1996). Our value is higher than the fractal dimension they determined for the penumbra. We have extended our study to include the umbra as well. Our values match well with the findings of a totally unconnected study on the fractal dimension of gold aggregates (Weitz *et al.*, 1985) and therefore we propose that the kinetic dynamics in the colloid system may be related to similar processes of compressible flow in the sunspot thus leading to similar morphology of these unrelated structures. Preliminary studies of the change of the fractal dimension for the umbra and penumbra temporally found that they are correlated and can indicate that the changes in complexity between the umbra and penumbra are linked.

**Acknowledgement:** The authors would like to thank the anonymous referee(s) for very valuable comments to improve the manuscript.